\documentclass[%
 reprint,
 amsmath,amssymb,
 aps,
]{revtex4-2}
\usepackage{xcolor}
\usepackage{hyperref}
\usepackage{graphicx}
\usepackage{dcolumn}
\usepackage{bm}

\graphicspath{{figs/}}

\begin{document}

\preprint{APS/123-QED}

\title{Information-theoretic language of proteinoid gels: Boolean gates and QR codes}


\author{Saksham Sharma}
\affiliation{Cambridge Centre for Physical Biology, Cambridge; Unconventional Computing Laboratory, UWE Bristol, UK}
\author{Adnan Mahmud}
\affiliation{Department of Chemical Engineering, Cambridge, Philippa Fawcett Drive, Cambridge CB3 0AS, UK; Zuse Institute Berlin, Takustraße 7 14195, Germany}
\author{Giuseppe Tarabella}
\affiliation{Institute of Materials for Electronic and Magnetism, National Research Council
(IMEM-CNR), Parma (Italy)}
\author{Panagiotis Mougoyannis}
\author{Andrew Adamatzky}
\affiliation{
Unconventional Computing Laboratory, UWE Bristol, UK
}%
\date{\today}

\begin{abstract}
Proteinoids are soft matter fluidic systems formed by heating and cooling of a mixture of different poly(amino acids) based polymers. Such systems are being shown previously to exhibit an oscillatory electrical activity inside, which can be treated as the core ingredient in their information carrying and transmission capacity. With an aim to build analog computers out of soft matter fluidic systems in future, this work attempts to invent a new information-theoretic language, in the form of two-dimensional Quick Response (QR) codes. This language is, effectively, a digital representation of the analog signals shown by the proteinoids. We use two different experimental techniques: (i) a voltage-sensitive dye and (ii) a pair of differential electrodes, to record the analog signals. The analog signals are digitally approximatied (synthesised) by sampling the analog signals into a series of discrete values, which are then converted into binary representations. We have shown the AND-OR-NOT-XOR-NOR-NAND-XNOR gate representation of the digitally sampled signal of proteinoids. Additional encoding schemes are applied to convert the binary code identified above to a two-dimensional QR code. As a result, the QR code becomes a digital, unique marker of a given proteinoid network. We show that it is possible to retrieve the analog signal from the QR code by scanning the QR code using a mobile phone. Our work shows that soft matter fluidic systems, such as proteinoids, can have a fundamental informatiom-theoretic language, unique to their internal information transmission properties (electrical activity in this case) - such a language can be made universal and accessible to everyone using 2D QR codes, which can digitally encode their internal properties and give an option to recover the original signal when required. On a more fundamental note, this study identifies the techniques of approximating \textit{continuum} properties of soft matter fluidic systems using a series representation of gates and QR codes, which are a piece-wise digital representation, and thus one step closer to programming the fluids using information-theoretic methods, as suggested almost a decade ago by \href{https://www.ams.org/journals/jams/2016-29-03/S0894-0347-2015-00838-4/S0894-0347-2015-00838-4.pdf}{Tao's fluid program}.

\end{abstract}

\maketitle
Proteinoids carry information in the form of oscillatory electrical (analog) signals. This oscillatory information can be interpreted as a core ingredient in architecting algorithms such that the proteinoid ensembles act as an \textit{analog computer} in solving varieties of computational tasks. Though not shown in proteinoids, there are many fluidic soft matter systems that have been shown to have such algorithmic capabilities. In 1950s, Ishiguro developed a machine to convert hydrographic and meteorological data into voltage and current signals in order to electronically model storm surges and tidal phenomenon, using wave-form generators and oscilloscopes \cite{ishiguro1959method}. \newline 

Water waves in a bucket was shown to solve the XOR problem and perform speech recognition using a machine called Liquid State Machine \cite{fernando2003pattern}. In another related work, Faraday waves observed on the surface of water were accessed as information-processing substrates by using them to design timers \cite{nakajima2015memory}. Recently, shallow water waves, given by Korteweg-de Vries (KdV) equation, are shown to have solutions that can encode information and aid in designing artificial neural networks of the form of echo state networks \cite{marcucci2023new, pierro2023optimization}. More recently, the nonlinear oscillation of gas bubbles inside water were shown to be integrable with the linear readout of an ESN algorithm, and thus, a bubble-based computing system can be used to forecast different chaotic time series \cite{maksymov2022neural}. Another experimental validation using solitary waves travelling on the surface of a flowing liquid film using reservoir computing systems was performed here \cite{maksymov2024physical}. \newline

\begin{figure*}[!tbp]
  \includegraphics[width=1.0\textwidth]{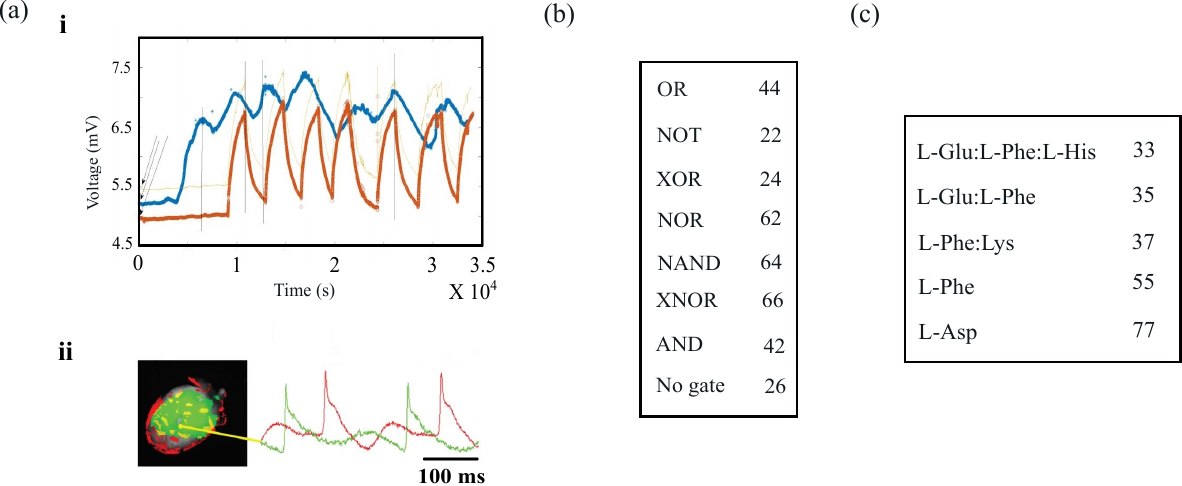}
  \caption{(a) Voltage versus time signal of the oscillatory electrical activity inside proteinoids measured using (i) differential electrodes and (ii) voltage sensitive dyes. (b) Table of the encoding scheme used for Boolean gates. (c) Table for the encoding scheme used for the proteinoid type.}
   \label{fig:fig1}
\end{figure*}
One crucial step in implementing such \textit{soft matter fluidic} algorithms is that the appropriate \textit{continuum} parameter in the problem concerned, is sampled at equidistant time instances. The prediction using reservoir computing (RC) algorithms, thereafter, is only \footnote{This can be seen as a physical version of the Church-Turing thesis: while the original Church-Turing thesis argues that an algorithm exists for what is effectively calculable using pen-and-paper method, in the context of physical problem, it can be interpreted as follows: the physical systems which are governed by underlying analytical (pen-and-paper) solutions can perform Turing-type computation of arbitrary precision.} possible if there is an analytical solution to the equation concerned that can be replaced by the reservoir computing network. As a result, this technique is still limited to physical problems where the underlying solution is known. The algorithmic capabilities of nonlinear waves, such as solitons and KdV waves, in the form of their Turing computability is rigorously established \cite{gracca2021computability}. However, this is not the case for the Navier-Stokes equations which describe dynamics of arbitrary soft matter fluidic systems, because of the lack of \textit{general} analytical solution \cite{sharma2022complexity} that can aid in Turing-computability and a built-in recipe for algorithm-generation capability \cite{sharma2022navier} for the problem concerned.\newline

By choosing to not much focus on what is fundamentally permissible and what is not, to meet the industry standards of analog computing, it might be beneficial to increase the power of soft fluidic computing algorithms. One way of doing this is dispersing microparticles inside the fluid. Few previously explored application in this direction are the microparticle-embedded fluids ~\cite{sason2017engineering, dhar2020self,pitingolo2018beyond}. When such model systems are actuated with electromagnetic radiations of suitable wavelength, the frequency of response to the actuation increases manifolds, making the system amenable to large computing power requirements. \newline

With the above-said pipeline in mind, we show in the present work, that arbitrary analog signals exhibited by the electrical property of poly(amino acid) ensembles can be used to construct digital fingerprint of the system in the format of QR codes, which can be used to access and retrieve back the original analog signals upon scanning. To the best of our knowledge, this has not been attempted before, and we believe that this marks the first step towards developing \textit{microparticle-embedded soft matter analog computers} in the future.\newline
 \begin{figure*}[!tbp]
  \includegraphics[width=1.0\textwidth]{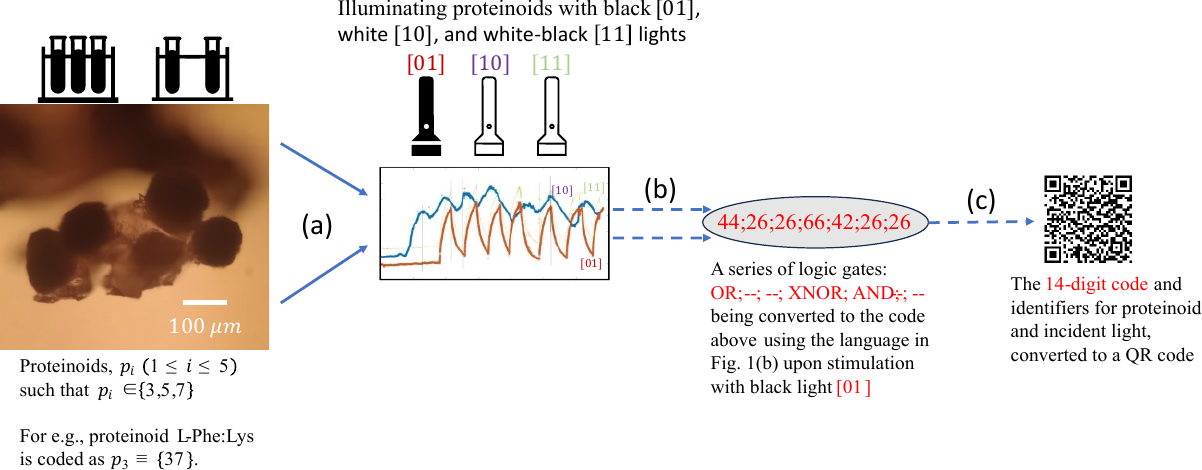}
  \caption{Projecting proteinoid solutions into QR codes (a) Subjecting the proteinoids in the test tubes to illumination through black, white, and black-white light to get the oscillatory reponse signal. (b) Converting the oscillatory analog response signal to a digital signal by extracting the logic gates from the analog signal. (c) Converting the series of logic gates in decimal format to a binary format which is then encoded in a QR code. The solid line represents an analog signal and dotted line represents a digital signal in processing.}
   \label{fig:fig1_5}
\end{figure*}
\section{Protocol for measuring electrical activity inside proteinoids}
\label{sec:protocol}

The proteinoids are prepared using the following apparatus \cite{sharma2023morphological}: powdered form of amino acids; Tri-block heater; 3-hole round bottom flasks; nitrogen gas; magnetic stirrer; cellophane membrane for dialysis; and a water bath. 1.5 grams each of L-Phenylalanine, L-Aspartic acid, L-Histidine, L-Glutamic acid and L-Lysine is mixed and heated in the 3-hole round bottom flask at 290$^{\circ}$C initially. The temperature is step-wise increased by 10$^{\circ}$ count, until fumes start to emerge out. The exhaust in the fumehood (with N$_{2}$ gas as the inlet) is started to release the fumes out. To begin with, the powder undergoes a color transformation from white to green, followed by a morphological transition resulting in the formation of a sequence of simmering microspheres. After ceasing the heating process, the remaining material solidifies and is subsequently extracted and allowed to cool for a duration of thirty minutes. The collected residue is then placed within the Slide-A-Lyzer mini dialysis apparatus, with 10,000 molecular weight cut-off and employs water as the dialysate. Dialysis is carried out continuously over a period of five days until the residue comprises minute ensembles of microspheres. The residue derived from the dialysis membrane is subjected to half an hour of heating in a vacuum oven, facilitating the evaporation of the dialysate (water) from the sample. Subsequently, the sample is scrutinized utilizing a transmission electron microscope. \newline

Subdermal iridium-coated stainless steel electrodes and potential-sensitive ANEP voltage-sensitive dyes were used to record the electrical activity. During the recording process, the data logger (ADC-24, Pico Technology, UK) operated at its maximum capacity (600 data points per second). This rate of data collection allowed for a comprehensive understanding of the electrical activity exhibited by the proteinoids. Furthermore, the data logger stored and saved the average value obtained from these measurements. This approach provided a concise representation of the recorded electrical activity, facilitating subsequent analysis and interpretation of the experimental results.

\section{Boolean gate representation of the electrical activity data}
\begin{figure*}[!tbp]
\includegraphics[width=0.8\textwidth]{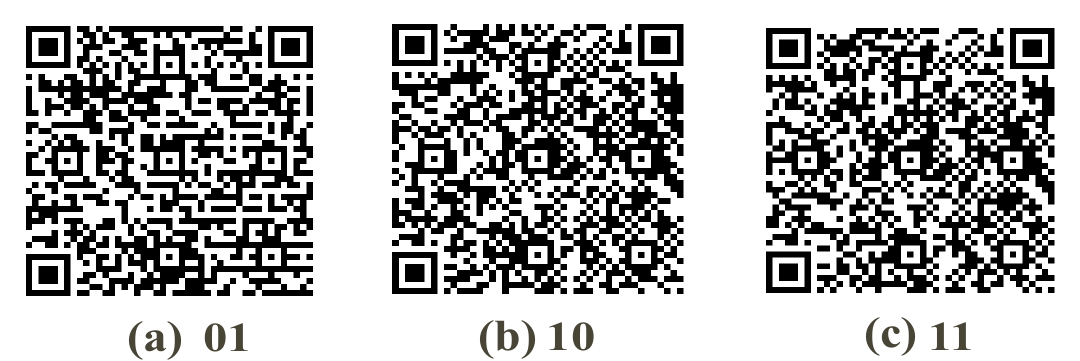}
\caption{QR codes generated for the proteinoids by characterising them based on the set of input lights: 01 (black and black-white); 10 (white and black light); 11 (black and white light).}
   \label{fig:fig2}
\end{figure*}

There are five different kinds of proteinoids which are being assessed in the data analysis shown in the present work. These proteinoids are, namely: L-Glu:L-Phe:L-His; L-Glu:L-Phe; L-Phe:L-Lys; and L-Asp. The electricaly activity recorded using the iridium-coated stainless steel electrodes and voltage-sensitive dyes identifies that the periods median of these proteinoids are distributed, fairly uniformly, at around 3500 seconds, except L-Phe which is at 2200 seconds. Subsequently, fast fourier transform (FFT) is performed to decomposed the time-varying signal into its frequency counterpart; further details of the analysis are available here \cite{mougkogiannis2023spiking}. A typical signal recorded is shown in Fig. \ref{fig:fig1}(a). The next step is to select a window of a period of 3000 seconds and identify underlying spikes in that window and rewrite it in the form of Boolean gates. The procedure on how to implement this has been previously discussed here \cite{adamatzky2019computing}. Once this procedure is successfully implemented, the spikes in the window are usually classified into one of the several Boolean gates enlisted in Fig. \ref{fig:fig2}(b). These Boolean gates are denoted by $g_{i} \, (1 \leq i \leq 8)$, such that, $g_{i} \in \{2,4,6\}$. For example, $g_{1} \equiv \textrm{OR gate}  \equiv 44$, $g_{2} \equiv \textrm{NOT gate} \equiv 22$, $g_{3} \equiv \textrm{XOR gate} \equiv 24$ and so on. It is to be carefully noted that if none of the Boolean gates in a given window, then the symbolic notation for ``No gate'' is 26. Similar to how the gates are named using the strings formed from symbols $\{2,4,6\}$, it is possible to encode the type of proteinoid using a different set of symbols. Let us denote proteinoids as $p_{i} \, (1 \leq i \leq 5)$, such that, $p_{i} \in \{3,5,7\}$. To give an example, $p_{1}$ denotes the proteinoid with composition L-Glu:L-Phe:L-His and is denoted by ``33'',  $p_{4}$ denotes proteinoid with composition L-Phe and the symbol ``55''.

\section{Boolean gate to QR code conversion of electrical activity}
In the previous section, we discussed the schemes adopted to encode different types of proteinoids and the corresponding Boolean gates. Using these schemes, in this section, we will characterise proteinoids and the signals hidden, and divide them in three categories:

\begin{figure*}[!tbp]
\includegraphics[width=0.8\textwidth]{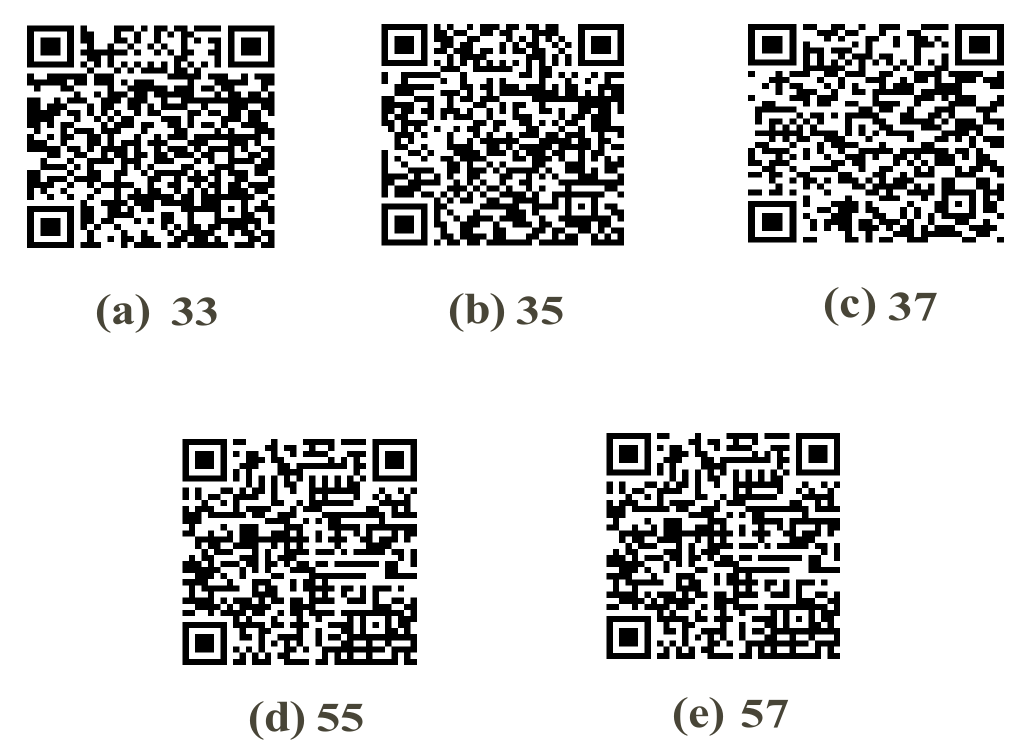}
\caption{QR codes generated for the proteinoids by characterising them based on the set of proteinoid types: 33 (L-Glu:L-Phe:L-His), 35 (L-Glu:L-Phe), 37  (L-Phe:Lys), 55 (L-Phe), and 57 (L-Asp).}
   \label{fig:fig3}
\end{figure*}

\begin{figure*}[!tbp]
\includegraphics[width=0.9\textwidth]{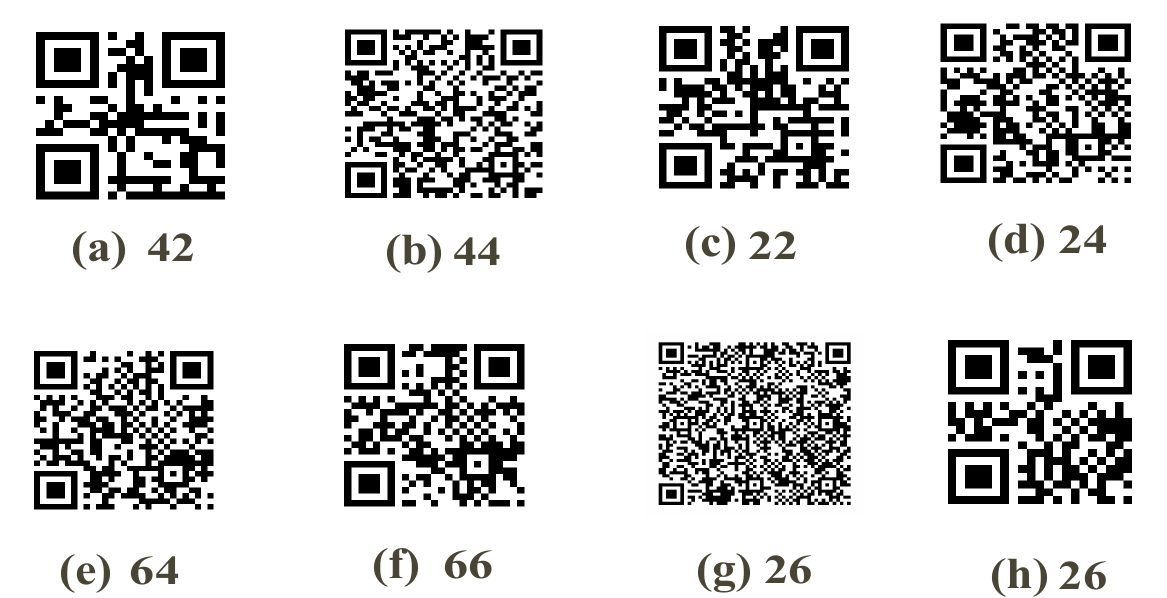}
\caption{QR codes generated for the proteinoids by characterising them based on the type of Boolean gates: 42 (AND), 44 (OR), 22  (NOT), 24 (XOR), 64 (NAND), 66 (XNOR), 62 (NOR), and 26 (no gate)}
   \label{fig:fig4}
\end{figure*}
\begin{enumerate}
    \item \textit{Characterisation based on the input lights:} 
    There are three different set of input lights that illuminate proteinoids so as to produce time-varying voltage signals. If the proteinoids are illuminated with black and black-white light periodically, it is encoded as ``01''. If the proteinoids, on the other hand, are illuminated with white and black light periodically, then it is encoded as ``10''. Lastly, when the periodic illumination is done using black and white light, then it is be encoded as ``11''. Based on these primary encoding schemes, the window carrying time-varying signal is converted to a series of Boolean gates for each of the proteinoid types. To given an example, the QR code in Fig. \ref{fig:fig2}(a) when converted in the decimal format reads as: $``334422642626242635442264262624263744262666\\42262655442664262624267744266426262426''$. While this may seem to be a large number difficult to decrypt, a close look reveals that the string begins with $p_{i} \in \{3,5,7\}$ followed by a series of eight Boolean gates $g_{i} \in \{2,4,6\}$, and again followed by a new proteinoid $p_{i+1}$ accompanied by the next seven Boolean gates next to it. Through the same procedure, the proteinoid types and corresponding Boolean gates are encoded in QR codes for the periodic light of the type $\{10\}$ and $\{11\}$.
    
    \item \textit{Characterisation based on proteinoid types:}
    It is also possible to characterise the signals in terms of the type of proteinoids. Given that there are five different kinds of proteinoids, $p_{i} \in \{3,5,7\}$, it means that there are five different kinds of QR codes possible to be encrypted, as shown in Fig. \ref{fig:fig3}. To given an example, the decimal format of the binary code encoded in Fig. \ref{fig:fig3}(a) is: $10442264262624261144226426262426014422642626\\2426$. It is apparent that the binary code starts with ``10'' which is the description of input light condition (white and light, periodically) followed by logic gates $g_{i}$. This is then followed by ``11'' which corresponds to periodic black and white illumination and again followed by series of logic gates, and so on. Using this procedure, the Boolean gates corresponding to the pattern of light illuminated is encoded for each proteinoid $p_{i} \in \{3,5,7\}$ in the QR code that is shown in Fig. \ref{fig:fig3}. 

    \item \textit{Characterisation based on type of Boolean gates:}
Lastly, it is possible to categorise the proteinoids based on the type of Boolean gates. Since there are eight type of logic gates, there are eight different type of QR codes shown in Fig. \ref{fig:fig4}. To take an example, consider Fig. \ref{fig:fig4}(a), where the binary code encoded, when converted into decimal format, yields: $10370137$. This means that when concerning AND gate, the proteinoid of the type $p_{3}$, i.e., L-Phe:Lys, exhibits such a gate when the lights are illuminated corresponding to ``10'' (white and black) and ``01'' (black and black-white) encoding.  
    
\end{enumerate}

\section{Projecting proteinoid solutions onto QR codes}
The process of projecting proteinoid solutions (a continuum matter) onto QR codes, is effectively, a pipeline for analog-to-digital (ADC) conversion. As shown in the schematic (Fig. \ref{fig:fig1_5}), this involves taking proteinoids $p_{i} (1 \leq i \leq 5)$ and illumating them with black, white, or white-black lights to yield an analog response which is discretised further to yield logic gates. By following the notation introduced in Fig. \ref{fig:fig1}, the sequence of logic gates, thus obtained, are converted to QR codes using a suitable code available at request. A web-interface to convert the string to a QR code (or vice-versa) is available in our \href{http://physicalcomputinglearning.pythonanywhere.com/}{\textcolor{blue}{website}}.

\section{Pipeline for digital encoding of soft matter fluidic systems}
Given that the oscillatory electrical activity inside the proteinoids is converted successfully into a piecewise discretised information (in the format of QR codes), it is an obvious next question to ask: what is the purpose of such a task? As originally alluded in the starting of this article, we aim to discard the \textit{continuum} description of properties of a gel (proteinoids in this case) and sample their discrete values in order to provide an information-theoretic description. This constitutes the first step, in a supposed pipeline presented in Fig. \ref{fig:fig5}, involving the measurement of continuum properties of soft matter fluidic systems and representing them in an analog fashion (such as, while recording in an oscilloscope). Once the \textit{measurement} is performed, the next step is to select the window and sample data from that window (after performing fourier transform and related data conversion techniques), in order to convert the analog signal to a piecewise data which can be further mapped from a decimal format to a binary format. This step is called \textit{discretisation}. In the present case, QR code formulation of discretised continuum data is an exemplary case to showcase this step. The obvious next step is to scan the QR code, which is indeed another form of measurement, or more precisely, an \textit{ante-measurement}, where the idea is to obtain the binary format of the continuum data back from the QR code description of continuum properties of soft matter fluidic systems. It is to be noted that in most of the cases where the data is not extraordinarily large, it is possible to retrieve the original \textit{discretised data} to 100\% efficiency. After the data is retrieved, it is possible through appropriate \textit{gluing techniques} to construct the original continuum level description of the soft matter system from the binary data. This \textit{retrieval step} is the most important and technique-dependent step that should be undertaken carefully. More on this will be discussed later.  

\section{Conclusion}
The present article presents a novel information-theoretic language of proteinoid gels, which are shown to have electrical oscillatory activity, amenable to a binary description using the language of Boolean gates and two-dimensional QR codes. We believe that the present work will guide the researchers in the field of proto-cognition, polymers, and microfluidics to consider QR codes and Boolean gates as a potential universal language of quantitative description of the behaviour of soft matter fluidic systems.

\section{Future directions and perspective}
One of the emerging perspective from this article is developing a pipeline (using idea of QR codes) to abridge the gap between continuum and discrete descriptions of soft matter fluidic systems. This is particularly useful when spiking activity of proteinoids is to be made universal and a unique marker of the proteinoid gel. The next step after such a representation is to regard the spiking activity using the typology of Izhikevich neurons. Such a characterisation will help us to throw some light on the ``neuroscience of proteinoids'', where the learning-cognitive behaviour of proteinoids is that of primal level, as often alluded with the term ``proto-cognition''.

\section*{Code}

The GUI used to encode and decode the QR codes for proteinoids is available \href{http://physicalcomputinglearning.pythonanywhere.com/}{\textcolor{red}{here}} and is written solely by A.M.

\bibliography{proteinoidbib1}

\end{document}